\documentclass{article}

\usepackage[english]{babel}

\usepackage[letterpaper,top=2cm,bottom=2cm,left=3cm,right=3cm,marginparwidth=1.75cm]{geometry}
\usepackage{cite} 
\usepackage{amsmath}
\usepackage{graphicx}
\usepackage[colorlinks=true, allcolors=blue]{hyperref}
\date{}

\begin{document}
\title{GQD-AdsNet: Graph Neural Networks Unlock Rapid Exploration of Transition Metal Adsorption on Graphene Quantum Dots}

\author{Lara Goncebat$^1$, Rodrigo Echeveste$^2$, Matías Gerard$^2$, Frederik Tielens$^3$,\\
Gustavo Belletti$^1$, Paola Quaino$^{1,*}$ \\
$^1$ Instituto de Química Aplicada del Litoral \\ IQAL (UNL-CONICET) Santa Fe, Argentina\\
$^2$ Instituto de Investigación en Señales, Sistemas e Inteligencia Computacional \\ sinc(i) (UNL-CONICET) Santa Fe, Argentina\\
$^3$ General Chemistry (ALGC) - Materials Modelling Group,\\ 
Vrije Universiteit Brussel (VUB), 1050 Brussel, Belgium\\
$^*$ Corresponding author. E-mail: pquaino@fiq.unl.edu.ar
} 
\maketitle

\begin{abstract}
In recent years, interest in single-atom catalysts supported on carbon-based structures has grown considerably due to their high catalytic activity and efficient uses of metal atoms. However, the design and characterization of these materials through first-principles calculations are computationally expensive, limiting the exploration of a large number of possible configurations. Here, we developed a framework based on graph neural networks (GNNs) to predict the adsorption energies of transition metals on graphene quantum dots (GQDs). The model was trained using data obtained from density functional theory calculations and achieved an R\textsuperscript{2} of 0.906 with an MAE of 0.101 eV, while reducing computational cost by roughly six orders of magnitude relative to DFT. This methodology provides an efficient tool for the accelerated screening and rational design of new catalysts based on carbon nanostructures.
\end{abstract}

\section*{Introduction}\label{sec1}

The increasing environmental degradation, climate change, and rapidly growing global energy demand have intensified the need for sustainable and efficient energy conversion technologies. In this context, electrochemical energy systems such as fuel cells, water electrolysis devices, and rechargeable metal–air batteries have emerged as promising alternatives for clean energy generation and storage \cite{faizan2026,wang2023}. The performance of these technologies strongly depends on the development of highly active, stable, and economically viable electrocatalysts.

Transition metals exhibit remarkable catalytic activity toward a broad range of electrochemical reactions. Nevertheless, their large-scale implementation is hindered by their high cost and limited natural abundance. To address these limitations, substantial efforts have been devoted to the development of single-atom catalysts and metal adatom-based systems, which maximize atomic utilization efficiency while preserving or even enhancing catalytic performance \cite{peng2018, araujo2024, urso2025} .

The catalytic performance and stability of single-atom systems are strongly influenced by the nature of the supporting material. Among the different candidates, carbon-based nanostructures have emerged as highly attractive supports owing to their excellent electrical conductivity, chemical stability, and tunable surface chemistry. These materials facilitate efficient charge transfer and promote the homogeneous dispersion of active sites, while simultaneously suppressing nanoparticle aggregation, sintering, and catalyst deactivation under electrochemical operating conditions \cite{hao2024}.

Within this family, graphene quantum dots (GQDs) constitute a particularly promising class of support materials. GQDs combine the exceptional electronic properties of graphene with finite-size and quantum confinement effects arising from their nanoscale dimensions \cite{lim2015}. In addition, their large surface-to-volume ratio, tunable electronic structure, and high density of chemically active edge sites provide favorable environments for strong metal–support interactions \cite{shen2012,li2013}. Experimental and theoretical studies have demonstrated that transition metal atoms anchored on GQDs can exhibit enhanced catalytic activity and stability for a variety of energy-related electrochemical reactions \cite{abdelsalam2018,ghosh2022,shashika2026,kumar2026,belletti2024}. In these systems, the adsorption energy of the metal adatom plays a central role, as it directly governs catalyst stability, resistance to sintering, and the electronic interactions responsible for catalytic activity.

Despite their promising properties, the rational exploration of metal–GQD systems remains computationally challenging. The structural diversity of graphene quantum dots, combined with the wide range of possible transition metal species and adsorption sites, generates a highly complex configurational landscape. Accurately evaluating adsorption energetics across this vast design space using Density Functional Theory (DFT) calculations is computationally demanding, making large-scale screening studies impractical.

Recent advances in machine learning (ML) have created new opportunities for accelerating materials discovery by enabling predictions with near first-principles accuracy at a fraction of the computational cost \cite{xie2018,nematov2025}. Among the different ML approaches, Graph Neural Networks (GNNs) have demonstrated exceptional capability for modeling atomistic systems  \cite{reiser2022}. These models are specifically designed to operate on graph-structured data, where node representations are iteratively updated through the aggregation and transformation of information exchanged with neighboring nodes according to the graph topology \cite{Scarselli09, Gilmer17}. Through successive message-passing operations, GNNs learn latent representations that capture both local atomic environments and long-range structural dependencies, enabling accurate prediction of physicochemical properties. In this context, the GQD-Metal system can be represented as a graph in which nodes correspond to atoms and edges encode chemical bonds or interatomic interactions. By propagating information through the graph, GNNs can effectively learn both pairwise and many-body interactions, allowing the extraction of structural and chemical features relevant to the prediction of material properties. Furthermore, GNNs naturally encode the local chemical environments and topological characteristics that govern adsorption phenomena. This capability makes GNNs particularly well suited for finite and non-periodic nanostructures such as graphene quantum dots, where adsorption properties are strongly influenced by local coordination environments, edge states, and structural heterogeneity.

While GNNs have achieved remarkable success in predicting properties of crystalline and periodic materials, comparatively fewer studies have addressed adsorption energetics on finite low-dimensional nanostructures, where edge effects, structural heterogeneity, and the lack of translational symmetry complicate the construction of transferable representations \cite{xie2018,reiser2022,ghanekar2022}. There is still, to the best of our knowledge, a lack of transferable graph-based frameworks for predicting adsorption energetics through a systematic and computationally efficient approach that enables the rapid screening of diverse transition metal–GQD configurations.

\begin{figure}
    \centering
    \includegraphics[width=1\linewidth]{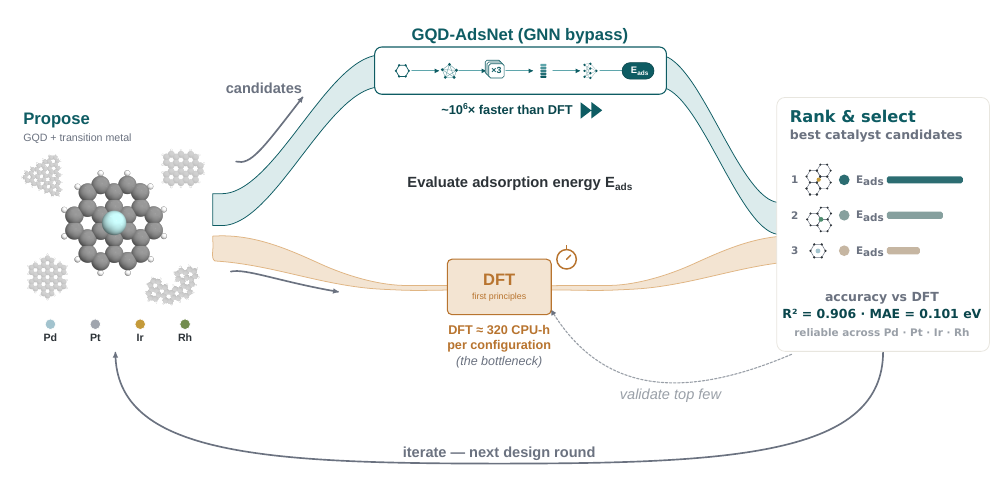}
    \caption{The schematic figure illustrates a gain up to a million-fold in speed when using GQD-AdsNet for compound screening versus traditional simulation methods}
    \label{conceptual}
\end{figure}

In this work, we present GQD-AdsNet (Graphene Quantum Dot Adsorption-Energy Network), a GNN framework for the rapid prediction of adsorption energies of transition metal adatoms on GQDs. The proposed approach enables efficient and low-cost exploration of complex adsorption configurations while maintaining high predictive accuracy relative to first-principles calculations. Figure \ref{conceptual} illustrates the computational bottleneck of DFT-based screening and the acceleration achieved through the proposed GNN-based prediction framework. Our framework establishes an effective route for the data-driven discovery of low-dimensional catalytic materials and highlights the potential of graph-based machine learning methods for accelerating the design of stable and efficient single-atom electrocatalysts.

\section*{Results and discussion}\label{sec2}

\subsection*{Systems}

\begin{figure}[h]
\centering
    \includegraphics[scale=0.35]{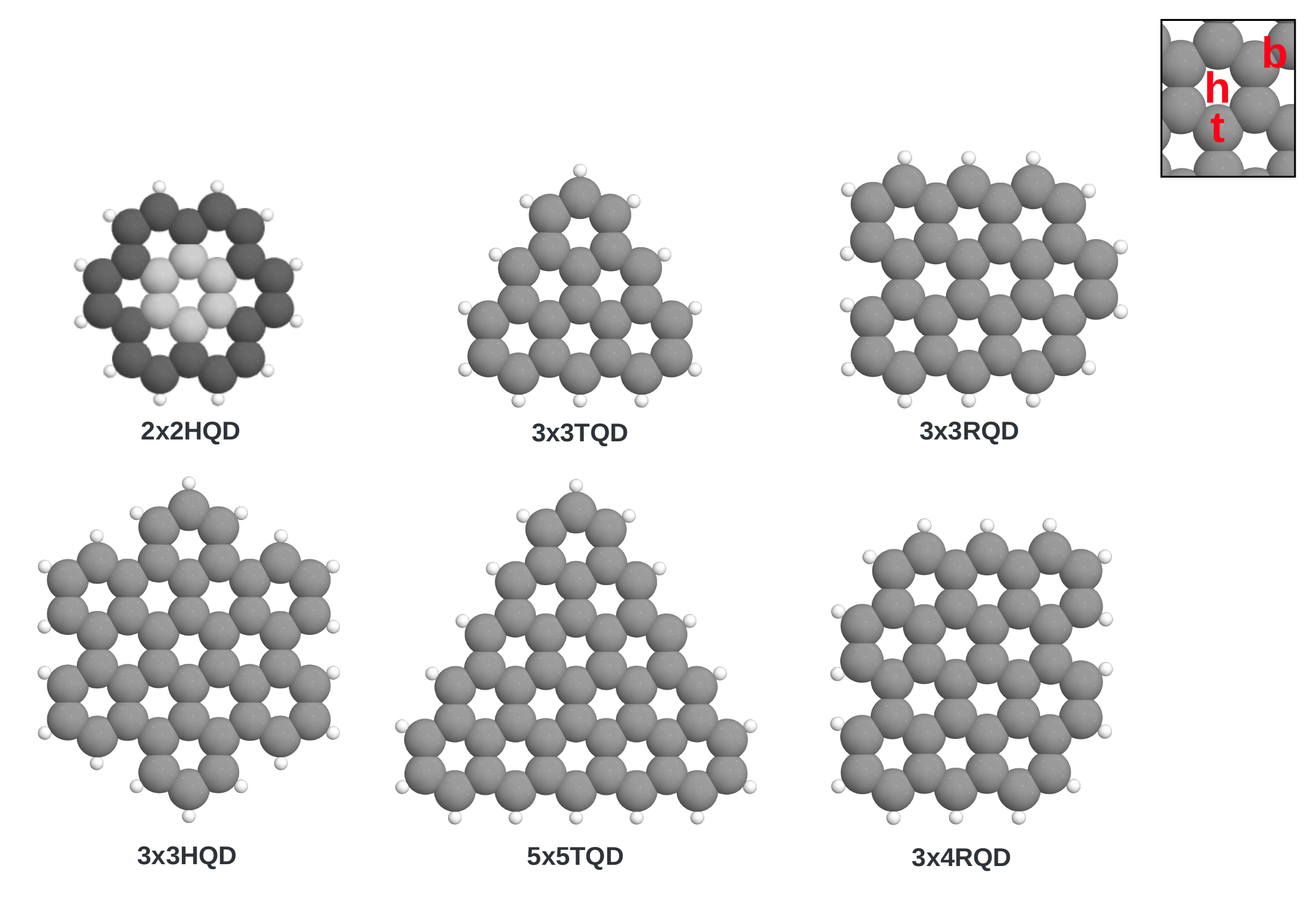}
    \caption{Graphene quantum dots: hexagonal quantum dots (HQD): 2x2HQD, 3x3HQD; triangulene quantum dots (TQD): 3x3TQD, 5x5TQD; rectangular quantum dots (RQD): 3x3RQD, 3x4RQD.  Carbon atoms are shown in gray, while hydrogen atoms are shown in white. The light and dark gray colors in the 2x2HQD indicate internal and external sites, respectively. The inset shows the high-symmetry adsorption sites: hollow (h), bridge (b), top (t).}
    \label{system}
\end{figure}

For the training and evaluation of the machine learning model, an original dataset composed of 491 graphene quantum dot–metal (GQD–M) adsorption systems was constructed, entirely generated through \textit{ab initio} calculations performed within the framework of the present work. Adsorption geometries were systematically constructed by placing transition-metal adatoms—palladium (Pd), platinum (Pt), iridium (Ir), and rhodium (Rh)—on all high-symmetry adsorption sites of six graphene quantum dots (GQDs) selected as reference systems; illustrated in Fig.~\ref{system}. 

Both internal and external adsorption sites were considered in order to capture variations in local coordination, symmetry, and edge-related effects. All studied adsorption sites, namely hollow (h), bridge (b), and top (t), are shown in the inset of Fig.~\ref{system}. The distinction between external and internal sites is indicated for a representative GQD in Fig.~\ref{system} using dark and light gray colors, respectively.

Each GQD–M configuration was treated as an independent system and fully relaxed using first-principles calculations to obtain reference adsorption energies ($E_{\mathrm{ads}}$), providing a diverse and physically consistent dataset for model training and evaluation.

\subsection*{GQD-AdsNet Performance}

\begin{figure}[h]
    \centering
    \includegraphics[scale=0.8]{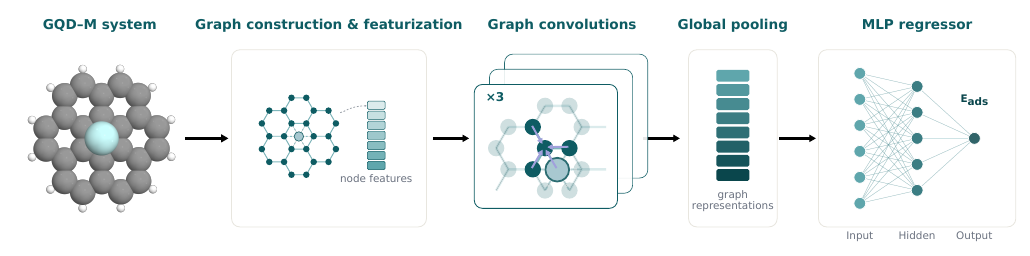}
    \caption{Architecture of the proposed GQD-AdsNet for adsorption-energy prediction. The graph representation of the GQD-M system is processed through three graph convolution layers followed by global max pooling and a multilayer perceptron (MLP) regressor.}
    \label{arq}
\end{figure}

In this work, we proposed GQD-AdsNet, a graph convolutional neural network consisting of three graph convolutional layers. Figure~\ref{arq} illustrates the overall architecture; additional details are provided in the Methods section.

\begin{figure}
    \centering
    \includegraphics[width=1\linewidth]{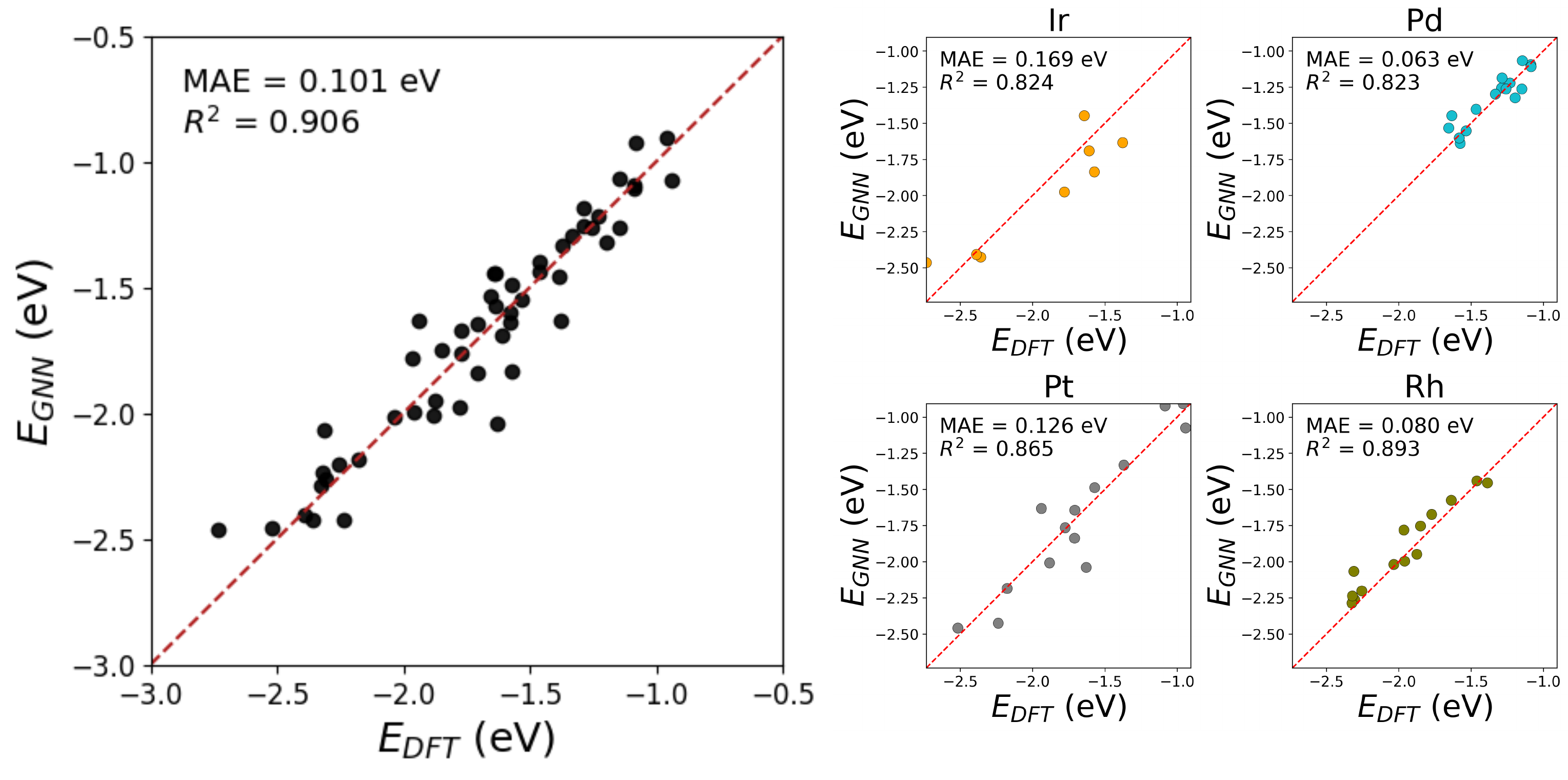}
    \caption{Adsorption energies predicted vs DFT calculated for the test set, categorized by transition metal (Ir: yellow, Pd: cyan, Pt: gray, Rh: green)}
    \label{test}
\end{figure}

The predictive performance of GQD-AdsNet was assessed on an independent test set using standard regression metrics. Fig.~\ref{test} shows the predicted adsorption energies as a function of the corresponding DFT-calculated values for the test configurations. The training and validation learning curves, together with the corresponding parity plots, are provided in the Supplementary Information (Figures S1 and S2).

The GQD-AdsNet achieves a coefficient of determination of $R^2 = 0.906$, with a root mean squared error (RMSE) of 0.134 eV and a mean absolute error (MAE) of 0.101 eV. The close agreement between the predicted and DFT-calculated adsorption energies demonstrates that the GQD-AdsNet successfully captures the local structural and chemical features governing metal adsorption on graphene quantum dots.

As an additional analysis, an ensemble predictor was constructed by averaging the predictions of the ten models obtained during the cross-validation procedure. Ensemble methods are known to improve predictive performance by reducing model variance and combining multiple hypotheses, often leading to better generalization capabilities \cite{Dietterich2000}. Consistent with this expectation, the ensemble model showed improved performance across all evaluation metrics. The MAE decreased to 0.097 eV, the RMSE  remained at 0.134 eV, and the R² increased to 0.907 Although the ensemble yielded slightly better performance metrics than the single model, the improvement was modest. This behavior suggests that the individual models converged to similar predictive functions, limiting the diversity required for larger ensemble gains. Therefore, the simplest single-model approach was selected as the final model, while the complete ensemble results are reported in the Supplementary Information.

The transferability of the model to previously unseen graphene quantum dot geometries was further evaluated using a leave-one-group-out scheme where the groups corresponded to the three GQD geometries considered in this work. The resulting MAE values were 0.133 eV for HQDs, 0.239 eV for TQDs, and 0.131 eV for RQDs. As expected, the predictive performance decreased compared to the model trained on all geometries. The results suggest that predictive performance is more sensitive to the exclusion of triangular geometries from the training set. This behavior is consistent with the more complex edge environments and distinctive electronic structure of triangular GQDs. A detailed description of this analysis and the complementary results are provided in the Supplementary Information.

To further investigate the model behavior, the predictive performance was analyzed separately for each metal species. Fig.~\ref{test} presents the parity plots grouped by metal. Although some variations in predictive accuracy are observed among different metals, the model maintains a consistently reliable performance across all systems. These differences may be attributed to variations in electronic structure and local coordination environments among the transition metals considered.

To evaluate the robustness of the model predictions, the distribution of prediction errors was analyzed in detail. The errors are centered around zero, indicating negligible systematic bias, with most deviations lying within approximately ±0.15 eV of the DFT reference values (Supplementary Information, Fig. S3). These results demonstrate the robustness of the GQD-AdsNet.

\subsection*{Generalization and Predictive Exploration}

To further evaluate the robustness and extrapolation capability of the trained GQD-AdsNet, predictions were performed on newly generated systems not included in the training dataset. Two complementary exploration strategies were considered: (i) systematic adsorption-site exploration on fixed graphene quantum dots and (ii) generation of new GQD structures.
Finally, both strategies were combined and integrated with the final trained model into a unified workflow for the rapid screening of adsorption configurations in previously unseen systems.

\subsubsection*{Adsorption-Site Exploration on Fixed GQDs}

\begin{figure}
    \centering
    \includegraphics[width=1\linewidth]{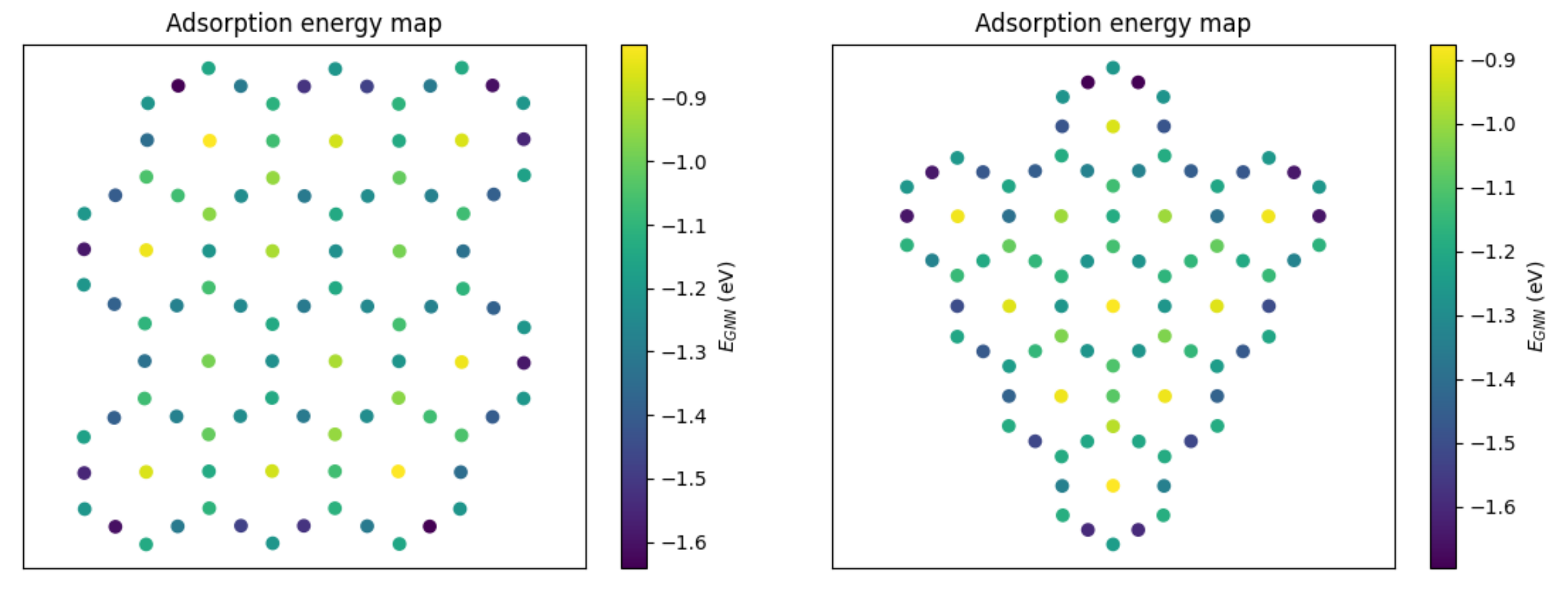}
    \caption{Adsorption energy maps of (left) the 3x4RQD-Pd system and (right) another GQD configuration with Pd adsorbed at all high-symmetry sites.}
    \label{maps_eads}
\end{figure}

In the first approach, a selected graphene quantum dot was used as the input structure, and adsorption configurations were systematically generated by automatically placing the metal atom on all high-symmetry adsorption sites. The corresponding graph representations were then evaluated using GQD-AdsNet. Fig.~\ref{maps_eads} (left) illustrates the predicted adsorption energies for all adsorption sites on a 3×4RQD. Due to the structural symmetry, several adsorption sites are equivalent; nevertheless, the model consistently reproduces the expected adsorption trends, yielding values in good agreement with those reported in the literature \cite{Manade2015, Hu2010}.

This procedure also enables the generation of adsorption energy maps for new GQD configurations that were not included in the dataset. As an example, the right panel of Fig.~\ref{maps_eads} shows the spatial distribution of $E_{\mathrm{ads}}$ values for a new proposed configuration.

More negative $E_{ads}$ values indicate stronger adsorption,
which is systematically predicted at external sites relative to interior positions, consistent with the lower coordination and enhanced electronic localization of edge atoms in finite graphene nanostructures \cite{belletti2024}. This indicates that the model captures physically meaningful adsorption trends beyond its training configurations.

\subsubsection*{Generation of New GQD Structures}

\begin{figure}
    \centering
    \includegraphics[width=0.9\linewidth]{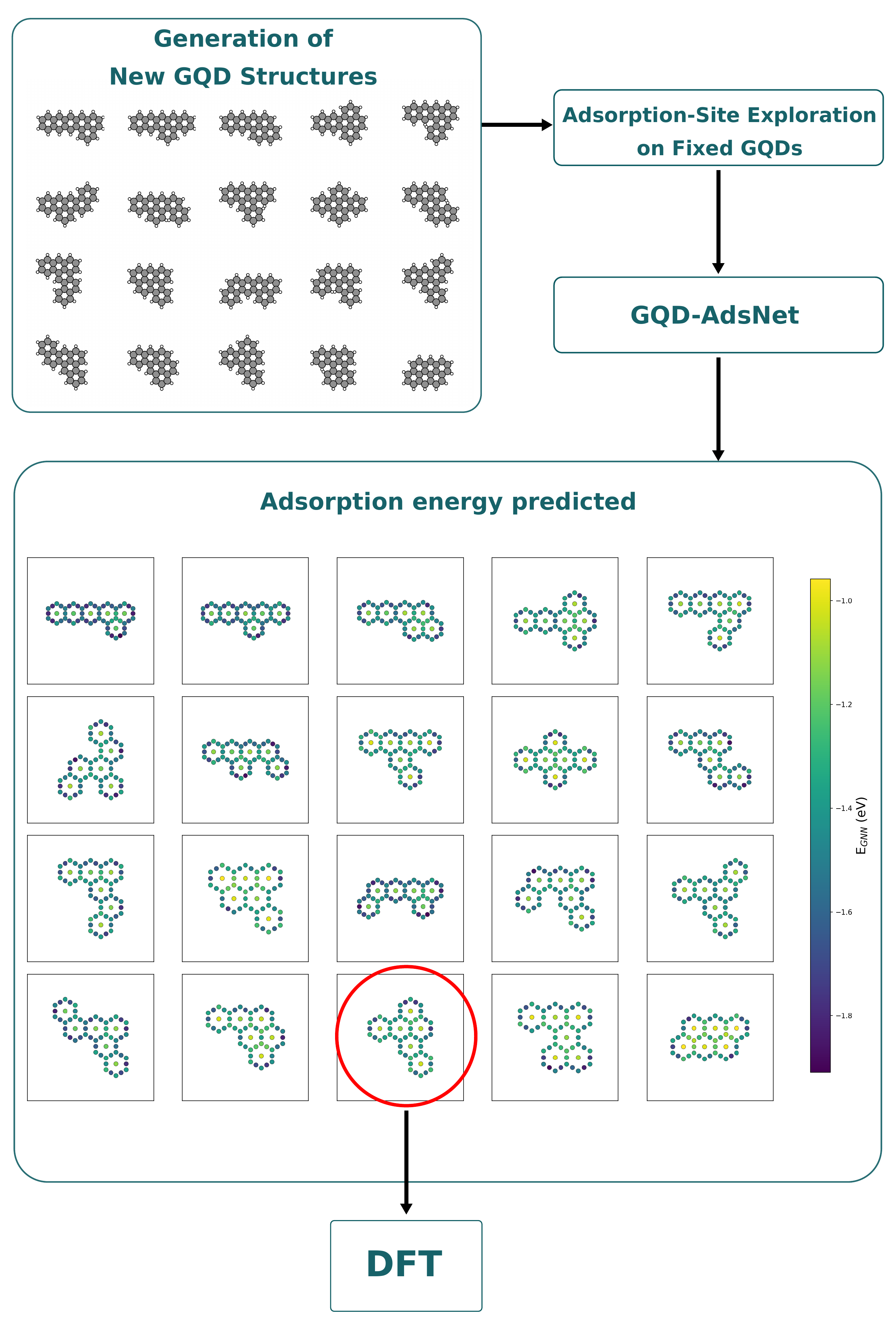}
    \caption{Illustration of the proposed workflow for the automated generation and screening of GQD adsorption systems. First, GQDs composed of six benzene units are generated automatically. Subsequently, all possible adsorption-sites configurations are constructed for a selected metal atom. The generated systems are then evaluated using GQD-AdsNet to predict adsorption energies, enabling the identification of the most promising catalytic configurations.}
    \label{workflow}
\end{figure}

In the second exploration strategy, entirely new graphene quantum dots were automatically generated from a predefined number of benzene units. The generation procedure produces all compatible non-linear GQD structures for a given size, enabling the exploration of previously unseen geometries and topologies. Combined with the Adsorption-Site Exploration and GQD-AdsNet, the workflow illustrated in Fig.~\ref{workflow} summarizes the overall strategy proposed in this work.

For each generated GQD, adsorption configurations were constructed and evaluated using the GQD-AdsNet. This automated workflow substantially expands the applicability of the proposed framework, enabling rapid exploration of adsorption properties in a broad range of graphene nanostructures without requiring additional DFT calculations.

Overall, these results demonstrate that the GQD-AdsNet is capable of extrapolating to new systems while preserving physically meaningful trends, supporting its potential use as an efficient tool for high-throughput computational screening of graphene-based catalytic materials.

\subsection*{Computational Cost Considerations}
An important aspect of this work is the comparison of computational cost between DFT and the GNN model. While a single DFT adsorption energy calculation may require several hours depending on the system size and computational resources, the GQD-AdsNet is capable of predicting adsorption energies for hundreds of generated configurations within seconds.
This difference becomes particularly significant when analyzing large configurational spaces, where the automated generation of graphene quantum dots and adsorption configurations can yield a large number of candidate systems. In such cases, the proposed workflow allows for the rapid and systematic exploration of this space, with a computational cost several orders of magnitude lower than that of conventional first-principles calculations. DFT calculations were performed using VASP on a 32-core CPU cluster (Intel(R) Xeon(R) Gold 6226R CPU, 2.90GHz) requiring approximately 320 CPU-h per adsorption configuration. In contrast, GNN predictions required only a fraction of a second ($\sim$0.002 s) per configuration on a single CPU core (Supplementary Information, Fig. S5). 

\section*{Conclusion}\label{sec13}

In this work, we developed a graph neural network framework to predict the adsorption energies of transition-metal adatoms on graphene quantum dots using structural and topological information derived from atomistic configurations. By combining first-principles calculations with graph-based machine learning, a diverse dataset of adsorption systems was constructed and used to train a predictive model capable of reproducing DFT adsorption energies with satisfactory accuracy.

The results demonstrate that graph neural networks can successfully capture the key physical factors governing adsorption phenomena in finite graphene nanostructures, even without the explicit inclusion of electronic descriptors obtained from DFT calculations. This finding suggests that local geometry and graph topology already encode a substantial portion of the relevant electronic information associated with adsorption processes.

In addition, the development of automated tools for generating graphene quantum dots and adsorption configurations enables the rapid exploration of previously unseen systems with minimal computational cost. This integrated framework provides a scalable strategy for accelerating theoretical studies and high-throughput screening of graphene-based catalytic materials.

Future work may focus on applying the proposed framework to catalytic reaction studies and to the prediction of additional physicochemical properties in graphene-based nanostructures.

\section*{Methods}\label{sec11}

\subsection*{Dataset and Graph Representation}
The dataset was constructed from the calculated GQD-M adsorption systems described in the Results and Discussion section, when each GQD-M system was represented as an undirected graph \( G=(V,E) \)\footnote{
A graph is formally defined as a pair G = (V, E), where V denotes a set of vertices and E a set of edges — each edge ($e_{v, w}$) encoding the relationship between two vertices v, w $\in$ V. When applied to molecular representation, this abstraction maps naturally onto chemical structure: atoms take the role of nodes, while the bonds connecting them are captured as edges, yielding an undirected graph.}, where nodes correspond to atoms and edges represent chemical bonds. Hydrogen atoms were excluded from the graph representation in order to reduce the complexity of the model and focus the learning process on the carbon lattice and adsorbed metal atom, which primarily govern the adsorption interaction.

Each node \( v_i \in V \) was characterized by seven descriptors encoding local structural and chemical information: atomic number, electron affinity, group, period, distance to the adsorbed metal atom, internal/external site indicator, and hydrogen-connectivity indicator. Additional details on the input features are provided in Table S1 of the Supplementary Information.

\subsection*{DFT Calculations}

All electronic structure calculations for the GQD-M systems were carried out within the framework of Density Functional Theory (DFT) using the Vienna Ab initio Simulation Package (VASP) \cite{vasp-a,vasp-b,vasp-c}. Valence electrons were described using a plane-wave basis set with a kinetic energy cutoff of 500 eV, while exchange–correlation effects were treated within the generalized gradient approximation (GGA) employing the Perdew–Burke–
Ernzerhof (PBE) functional \cite{pbe-1,pbe-2}. Long-range dispersion interactions were included through the DFT-D3 correction scheme \cite{d3-a,d3-b}, and spin polarization was taken into account in all calculations.

Due to the finite size of the systems, Brillouin-zone sampling was restricted to the $\Gamma$-point. Both GQD and GQD–M systems were fully optimized by relaxing the atomic positions using the conjugate-gradient method until the residual forces converged below 0.01 eV~\AA$^{-1}$.

The adsorption energy of a metal atom (M) on each GQD, considering all adsorption sites shown in Fig.~\ref{system}, was computed as shown in eq.~\ref{eq.1}:

\begin{equation}
E_{\mathrm{ads}} = E_{\mathrm{GQD-M}} - E_{\mathrm{GQD}} - E_{\mathrm{M}},
\label{eq.1}
\end{equation}

\noindent where $E_{\mathrm{GQD-M}}$ is the total energy of the adsorbed system, $E_{\mathrm{GQD}}$ is the energy of the isolated graphene quantum dot  in its equilibrium geometry and $E_{\mathrm{M}}$ corresponds to the isolated metal atom in the gas phase.

\subsection*{GQD-AdsNet Architecture}

The GQD-AdsNet is based on a Graph Convolutional Neural Network (GCNN) architecture implemented using PyTorch and PyTorch Geometric \cite{pytorch, pyg}. The convolutional layers iteratively update each node vector $v_i$ by aggregating information from neighboring atoms and bonds through nonlinear graph convolution functions.

\begin{equation}
    v_i^{t+1} = Conv (v_i^t, v_j^t, u_{(i,j)}).
\end{equation}

After three convolutions,  the network automatically learns the feature vector of each node by iteratively aggregating information from its local environment. Graph convolutions do not reduce the dimensionality of the graph. Namely, the output of each convolutional layer "lives" in a graph with the same nodes as the input. Predictions over a complete graph hence require matching the dimensionality of the last convolution layer, to that of the predicted quantity. This is achieved using a pooling function followed by a set of fully connected layers. The pooling function combines the atomic feature vectors to reduce the dimensionality and generate a single vector $v_p$, or "fingerprint", that captures the overall characteristics of each system.
\begin{equation}
    v_p = Pool (v_0^{(0)},v_1^{(0)},...,v_N^{(0)},...,v_N^{(R)}).
\end{equation}

The pooled vector $v_p$ is then fed into a multilayer perceptron (MLP) regression head consisting of two fully connected layers separated by a ReLU  (Rectified Linear Unit) nonlinear activation function. The output layer produces a scalar value corresponding to the predicted adsorption energy. The overall workflow of the model is shown in Fig.~\ref{arq}.

\subsection*{Model Training and Evaluation}

The dataset was first divided into training/validation (90\%) and test (10\%) subsets using a stratified split based on metal identity. The test set was held out throughout model development and was not used for training or hyperparameter optimization.

Feature normalization was performed using statistics computed exclusively from the training data to avoid data leakage. The model was trained in a supervised manner by minimizing the mean squared error (MSE) between predicted and DFT-calculated adsorption energies using the Adam optimizer. Early stopping based on validation loss was employed to prevent overfitting.

Final model performance was assessed on the independent test set using the root mean squared error (RMSE), mean absolute error (MAE), and coefficient of determination ($R^2$).

\subsection*{Hyperparameter optimization}

The hyperparameter optimization of the GQD-AdsNet was performed using metal-stratified K-fold cross-validation. The optimal configuration was selected based on validation performance. The final model consists of three graph convolution layers with 64 hidden features, an initial learning rate ($1\times10^{-3}$) and a batch size of 8. In addition, different global pooling operations (max, add and mean) were evaluated. Among them, global max pooling demonstrated the best predictive performance on the validation set and was therefore adopted to obtain graph-level representations.

\subsection*{Data availability}
 The dataset associated with this work is publicly available through Zenodo 

\noindent (https://doi.org/10.5281/zenodo.21421132) and on Github (https://github.com/litoralchem/GQD-
AdsNet)

\subsection*{Code availability}
The source code developed for this work is publicly available at GitHub

\noindent (https://github.com/litoralchem/GQD-AdsNet). 
An archived version of the
software is available through Zenodo (DOI: https://doi.org/10.5281/zenodo.21421132).

\subsection*{Acknowledgements}
Financial support by CONICET and Universidad Nacional
del Litoral is gratefully acknowledged. GB, LG, and PQ thanks CAID
85520240100069LI and PIP 11220210100127CO for support. The present work used
computational resources of the Pirayu Cluster, acquired with funds from the Santa
Fe Science, Technology and Innovation Agency (ASACTEI, Grant 00010-18-2014);
and from clusters within CCAD-UNC, which are part of SNCAD-MinCyT, Argentina.
The authors also gratefully acknowledge the SICyT and the administrative staff
for providing computational resources on Clementina XXI under project PISCA-65
(2025).

\end{document}